# Proposal for compact solid-state III-V single-plasmon sources

C. H. Gan[1,3], J.P. Hugonin[1] and P. Lalanne[1,2]

[1] *Laboratoire Charles Fabry, Univ Paris-Sud, CNRS, Institut d'Optique, Campus Polytechnique, 91127 Palaiseau Cedex, France*

[2] *Laboratoire Photonique, Numérique et Nanosciences, Université Bordeaux 1, CNRS, Institut d'Optique, 33405 Talence Cedex, France*

[3] *Electronics and Photonics Department, A*STAR Institute of High Performance Computing, 138632, Singapore*

**ABSTRACT**

We propose a compact single-plasmon source operating at near-infrared wavelengths on an integrated III-V semiconductor platform, with a thin ridge waveguide serving as the plasmon channel. By attaching an ultra-small cavity to the channel, it is shown that both the plasmon generation efficiency (β) and the spontaneous-decay rate into the channel can be significantly enhanced. An analytical model derived with the Lorentz reciprocity theorem captures the main physics involved in the design of the source and yields results in good agreement with fully-vectorial simulations of the device. At resonance, it is predicted that the ultra-small cavity increases the β-factor by 70% and boosts the spontaneous decay rate by a factor 20. The proposed design could pave the way towards integrated and scalable plasmonic quantum networks. Comparison of the present design with other fully-dielectric competing approaches is addressed.

## I. INTRODUCTION

Surface plasmon polaritons (SPPs), which are formed by a coherent superposition of electron oscillations and electromagnetic fields confined at conductor–dielectric interfaces, play an important role in subwavelength optical systems [1]. At energies close to the resonance frequency, they can achieve small modal wavelengths and large local electromagnetic field enhancements. Thus it seems natural to explore the potential of plasmonics not only to reduce the size of optoelectronic devices [2], but also to facilitate the coupling of light to the nanoscale, including quantum dots (QDs) or individual molecules, with the ultimate goal of achieving scalable quantum computers [3-8]. To this end, a key feature is the development of compact sources that are able to efficiently generate tightly-confined SPPs.

The predicament is the inevitable trade-off between localization and ohmic losses that become significant when the SPP electromagnetic energy is mainly stored inside the metal. Indeed, very large SPP-generation efficiencies are obtained whenever the transverse SPP-mode area is below the diffraction limit but, ineluctably, the SPP propagation distance $L$ becomes small. For instance, large generation efficiencies (~0.8-0.9) are obtained in [8] using V-groove plasmonic channels at visible frequency, but $L$ is only weakly larger than the operating wavelength. This fundamental trade-off between the coupling efficiency and incurred dissipative loss puts severe limitations on the integration and scalability of plasmonic integrated circuits [1] or quantum networks [8]. To design an optimal system, one needs to search for the right balance [9-11].

In the present work, we tackle the important problem of designing compact single-plasmon sources at near-infrared frequencies. The source is designed to be fully-compatible with the mature semiconductor III-V AlGaAs self-assembled QD technology, which represents a suitable platform for integrating optical quantum networks [12]. Bearing in mind the above-mentioned trade-off, we consider plasmonic waveguides with an effective mode area (see Fig. 1a) that is substantially smaller than those of dielectric channel counterparts, while maintaining a reasonably long propagation distance ($L$ ~15λ). We show that, by attaching an ultra-small (< 0.001 µm$^3$) cavity to the SPP waveguide, it is possible both to enhance the source efficiency, defined as the fraction of collected plasmons per trigger, and to increase the decay rate into the plasmonic channel. In addition to practical issues related to the compactness, ease of integration and scalability, the efficiency and the decay rate into the plasmon channel are key performance measures for single-plasmon sources; large efficiencies are required to maintain a strong degree of interaction in a quantum network, and high rates are desirable for high processing speeds and for reducing decoherence effects, which may undermine the plasmonic quantum interference capability of indistinguishable or entangled plasmons [13]. Overall, we believe that the present approach represents a good compromise between transversal confinement, coherence length, efficiency, and decoherence immunity.

## II. PLASMONIC CHANNEL

We start by considering the spontaneous emission of a self-assembled QD embedded in a thin GaAs ridge channel of width $d$ and thickness $t$ deposited on a gold substrate (see Fig. 1(a)). In our calculations, the refractive index of gold follows from tabulated data [14], and we assume that the self-assembled QD emitting at $\lambda_0 = 950$ nm ($k_0=2\pi/\lambda_0$) has an in-plane polarization (perpendicular to the growth axis) with a linear polarization $p_z e^{-i\omega t}$ along the propagation direction of the SPP-channel. The QD normalized total decay rate is given by $P' = \Gamma/\Gamma_0$ (the Purcell factor), where $\Gamma = 2|p_z|^2 \omega^2 \text{Im}[\mathbb{G}_{zz}(\mathbf{r'}, \mathbf{r'}; \omega)] \hbar\varepsilon_0 c^2$, and is calculated with a 3D fully-vectorial Fourier modal method [15,16]. Here, $\mathbb{G}_{zz}(\mathbf{r}, \mathbf{r'}; \omega)$ is the electric field Green tensor describing the field response at $\mathbf{r}$ of an oscillating dipole positioned at $\mathbf{r'}$, and $\Gamma_0$ is the SE rate in the bulk semiconductor. To avoid excessive quenching observed for QDs in the close proximity of metals, we further assume that the QD is located 20 nm above the gold surface [17]. Note that this assumption is legitimate in the present case, since the modulus of the $z$-component of the SPP-mode electric field is maximum at the top GaAs/air interface, see Fig. 1(a). Due to the proximity of the metal film, the SPP field is highly localized especially along the $y$-direction despite the very small thickness ($t = 30$ nm) of the GaAs waveguide. Figures 1(b) and 1(c) show the Purcell factor $P'$ and the fraction $\beta'$ of the spontaneous decay that is directly coupled to the two contra-directional SPP channels. Four tiny ridge thicknesses, 25 nm < $t$ < 40 nm, are investigated. $P'$ decreases for small widths ($d \lesssim 500$ nm), and rapidly tends towards an asymptotical value for $d > 1$ µm. Meanwhile, $\beta'$ decreases due to the increase in the SPP-mode effective surface area, $\beta' \sim d^{-1}$ as $d \to \infty$. With Fig. 1(d) we show the SPP propagation length $L$ as a function of the ridge width, and we illustrate the inevitable trade-off between localization and loss. Indeed a close inspection to Figs. 1(b)-1(d) reveals that decreasing either $t$ or $d$ to enlarge the SPP propagation length systematically lowers the QD decay rate $P'$ and the SPP excitation efficiency $\beta'$. For the ensuing investigations, the dimensions of the waveguide is taken to be $t = 30$ nm and $d = 500$ nm, in order to achieve a good compromise between large decay rate ($P' = 0.82$), high coupling efficiency ($\beta' = 20\%$), and relatively long interaction distances ($L \sim 10$ µm).

### III. DESIGN OF THE SINGLE-PLASMON SOURCE

The SPP channel waveguide with the attached nanocavity placed near to the QD is sketched in Fig. 2(a). The nanocavity, a nanogroove filled with a dielectric (refractive index 1.5), has lateral dimensions $w_x$, $w_z$, and depth $h$. The QD decay mechanism can be decomposed in two successive steps whereby the QD first excites the cavity mode, which in turn decays into the SPP channels. To quantify the performance of the single plasmon source *in the presence of the nanogroove*, let us define the following parameters that are central to the analysis

- $P \equiv$ total spontaneous decay rate normalized to $\Gamma_0$,
- $\beta \equiv$ plasmon generation efficiency, i.e. fraction of the spontaneous decay that is directly coupled to the two contra-directional SPP channels,
- $\gamma \equiv$ coupling coefficient between the QD and the downward traveling $TE_{01}$ mode in the nanogroove, see Fig. 2(c1).
- $\alpha \equiv$ coupling coefficient between the upward traveling $TE_{01}$ mode in the nanogroove and each SPP-ridge mode, see Fig. 2(b).

Note that no primed notation is used for the definition of these parameters to differentiate them with those introduced in the previous section in the absence of the nanogroove (for the waveguide geometry alone).

**Coupled-wave model.** To better evidence the electrodynamics of the system and to help the design, we first develop a coupled-wave (CW) model assuming that the fundamental $TE_{01}$ mode of the groove is the dominant mode bouncing back and forth in the groove (this assumption will be tested against fully vectorial calculations hereafter). Figure 2(b) shows the main physical quantities of the model. The near-field coupling coefficient $\gamma$ between the QD and the downward-traveling $TE_{01}$ mode in the nanogroove is obtained by applying the Lorentz reciprocity theorem [18], $\gamma = -\mathbf{E}(\mathbf{r}_0) \cdot \mathbf{p}_z/N_T$, where $N_T$ is a normalization constant of the $TE_{01}$ mode, and $\mathbf{E}(\mathbf{r}_0)$ is the electric field scattered at the QD position $\mathbf{r}_0$ by the upward-traveling $TE_{01}$ mode as in Fig. 2(c2). Details concerning the normalization can be found in [15]. A unique strength of the model is its ability to allow for an analytical treatment of the QD position $\mathbf{r}_0$, since the sole knowledge of the scattered field $\mathbf{E}(\mathbf{r})$ in Fig. 2(c2) allows us to derive $\gamma$ for any $\mathbf{r}_0$. More importantly, in comparison with earlier classical works on Fabry-Perot models dealing with emitters located inside cavities [19,20], the present approach is more general as it encompasses the important case of nanoantennas for which the emitter may be actually located outside the cavity [6,7] in the near field of the resonance mode. Note that such extern is also important for electrical pumping. The nanogroove resonance is described classically by the coefficients $A$ and $B$, which represent the amplitudes of the downward and upward $TE_{01}$ modes in the nanogroove; we further denote by $r_h$ and $r_b$ the modal reflection coefficients at the top and bottom facets of the groove, see Fig. 2(b). Using coupled-wave equations, one may obtain

$$A = \gamma/(1 - r_h r_b u^2) \text{ and } B = A r_b u^2, \qquad (1)$$

with $u = \exp(ik_0 nh)$, and $n$ the effective index of the $TE_{01}$ mode. The nanocavity quality factor is given by

$$Q = k_0 \, \text{Re}(n_g) \, L_{eff}/(1 - |r_h r_b u^2|), \qquad (2)$$

where $n_g = d(n\omega)/d\omega$ is the group index of the $TE_{01}$ mode, and $L_{eff}$ is the effective cavity length that includes the penetration lengths upon reflection at the two facets [21]. For subwavelength nanogroove depths, $|u^2| \sim 1$, and since $|r_b|^2$ ($\approx 0.97$ nearly independently of $w_z$) is also close to unity, the $Q$-factor (and hence the Purcell factor) is mainly limited by the scattering losses at the top facet of the groove, $|r_h|^2$ being large only for very small values of $w_z$ (see Fig. 3(b)).

To further analyze the efficiency of the plasmon source, we consider the coupling coefficient $\alpha$ between the nanogroove $TE_{01}$ and the SPP-ridge modes. To calculate $\alpha$, we use the mode orthogonality of translational-invariant lossy waveguide [22], which yields $\alpha = \oiint [\mathbf{E}(\mathbf{r}) \times \mathbf{H}_{SPP} - \mathbf{E}_{SPP} \times \mathbf{H}(\mathbf{r})] \cdot \mathbf{z}\, dS$, where the integral is performed over any transverse cross-sectional surface of the SPP channel, [$\mathbf{E}(\mathbf{r})$, $\mathbf{H}(\mathbf{r})$] are the electromagnetic fields scattered by the $TE_{01}$ mode as shown in Fig. 2(c2), and [$\mathbf{E}_{SPP}$, $\mathbf{H}_{SPP}$] are the normalized modal field components of the SPP mode.

The two-step emission mechanism neglects all the decay channels other than the cavity one. As shown in Fig. 2(b), the QD may directly excite the SPP channels or may directly radiate into free space. It is reasonable to anticipate that, except when the QD emission line is largely detuned from the nanocavity resonance frequency, the QD emission decay is predominantly funneled into the cavity and other decay channels become negligible. With this hypothesis, very simple analytical expressions for the main performance measures of the source are obtained. For instance, the normalized spontaneous decay-rate $\Gamma_A$ into ohmic losses in the nanocavity, the normalized spontaneous emission-rate $\Gamma_R$ into free-space photons, and the desired spontaneous decay rate into the SPP channels $\Gamma_{SP}$ read as $\Gamma_A^{(CW)} = (|A|^2 - |B|^2)/\Gamma_0$, $\Gamma_R^{(CW)} = |B|^2 (1 - |r_h|^2 - 2|\alpha|^2)/\Gamma_0$, and $\Gamma_{SP}^{(CW)} = 2|B\alpha|^2/\Gamma_0$, respectively, where the superscript (CW) is used to emphasize that the analytical expressions are only approximate and are obtained with the CW model. Thus the total decay rate is given by

$$P^{(CW)} = |A|^2 (1-|r_h r_b u^2|)/\Gamma_0, \quad (3)$$

and the efficiency $\beta^{(CW)} = \Gamma_{SP}^{(CW)} / P^{(CW)}$ of the single-plasmon source is

$$\beta^{(CW)} = 2|r_b u^2 \alpha|^2 / (1-|r_h r_b u^2|), \quad (4)$$

where the factor 2 accounts for the two excited channels.

To check the validity of the CW model, we have performed 3D fully-vectorial calculation using the aperiodic-Fourier modal method [15,16]. In comparing to the numerical results, it will be seen that the CW model is highly accurate even if it solely considers the fundamental $TE_{01}$ mode. In fact, the higher-order propagating modes mainly contribute to increase absorption loss and become relevant only for wavelengths significantly smaller than $\lambda_0$.

Figures 3(a) to 3(c) show the impact of the groove width $w_z$ on the scattering coefficients of the CW model; the $TE_{01}$ reflectance $|r_b|^2$ of the bottom facet being close to 1 is not shown. Three values of the groove length, $w_x = 250, 300,$ and 350 nm, are considered. Calculations (not shown) reveal that the coupling-coefficient $\gamma$ is nearly independent of the groove length $w_x$, but strongly depends on the groove width $w_z$. Starting from $w_z = 0$ ($\gamma = 0$), the coupled intensity $|\gamma|^2$ rapidly increases to a maximum for $w_z \approx 20$ nm, and then decreases monotonically as $w_z$ further increases. As seen from Fig. 3(a), for large $w_z \sim 200$ nm, roughly half of the total spontaneous decay is coupled to the $TE_{01}$ mode in the absence of feedback from the bottom facet, but for small widths $w_z \leq 20$ nm, the coupling is less efficient as the decay is mainly directed into free-space photons. As will be shown shortly, the cavity strongly influences the repartition of the QD decay into the different available channels. As shown in Fig. 3(b), the reflectance of the top facet remains large, $0.75 < |r_h|^2 < 0.95$ but we note that the reflectance strongly depends on the groove length, especially for the investigated lengths that are approximately equal to half of the wavelength in the dielectric material (cutoff for the x-direction). The coupling strength $|\alpha|^2$ between the $TE_{01}$ mode and each SPP-channel is shown in Fig. 3(c). Indeed, $|\alpha|^2$ and $|r_h|^2$ are not independent; we additionally note that $2|\alpha|^2$ remains slightly smaller than $1-2|\alpha|^2-|r_h|^2$, for all $w_z$'s, suggesting that the detrimental radiation into free space is not negligible.

A resonance of the nanocavity is achieved whenever the Fabry-Perot condition $2k_0\text{Re}(n)h + \arg(r_h r_b) = 2m\pi$ is satisfied ($m$ being an integer). The nanogroove depths $h$ corresponding to resonance at $\lambda_0 = 950$ nm are shown in the insets of Fig. 3(d). It is worthwhile noting that for $m = 0$, $h$ is virtually constant and very small ($h \approx 20$ nm for $w_z > 50$ nm). The Purcell factor $P$ and the SPP coupling efficiency $\beta$ are respectively shown in Figs. 3(d) and 3(e) for $w_x = 300$ nm. The solid curves represent the CW predictions of Eqs. (3) and (4) obtained for $m = 0$ or 1. Except for an unexpected enhancement near $w_z = 25$ nm for the case $m = 1$, the model predictions quantitatively agree with fully-vectorial computational results shown with dots or squares and obtained with the fully-vectorial aperiodic-Fourier modal method. Thus we may use the CW model with confidence for the following analysis.

**Impact of geometrical parameters.** For instance, the model well explains the complex dependence of the total spontaneous decay rate $P$ with $w_z$ with a large peak value for tiny nanogrooves. For small widths, $P$ is largely impacted by the coupling efficiency $|\gamma|^2$, which rapidly vanishes as $w_z \to 0$. At large widths, $P$ falls down because of the decrease of the top facet reflectivity $|r_h|^2$ and of the increase of the $TE_{01}$ mode effective cross-section area, which result in a decrease of the cavity Q and in an increase of the cavity-mode volume. The sharp enhancement for $w_z = 25$ nm and for $m = 1$ is not captured in the CW model as it arises from the resonance of a higher-order nanogroove mode (other than $TE_{01}$) that is not considered in the CW model [23].

Figure 3(e) shows that the dependence of the β-factor is also well-explained with the CW model. Except for the unimportant sudden increase of β at very small widths, (β → β' as $w_z \to 0$), the general trend is a progressive decrease of β as $w_z$ is reduced, unfortunately leading to dichotomic behaviors between β and $P$. For small nanogroove widths, $|r_h|^2 \approx 1$ and the QD decay-rate is high with a peak value ~ 40 times larger than the Purcell factor $P$' in the absence of cavity. However, the energy stored in the cavity is mainly lost as heat and the SPP-source efficiency is small. Reversely for large nanogroove widths, $|r_h|^2$ is lowered and out-coupling mechanisms are promoted; a large fraction of the stored energy is coupled out either as free-space photons or as SPPs.

## IV. PERFORMANCE OF THE SOURCE

Figure 4 summarizes the main useful results for the single-plasmon source. It shows the performance of the source for an intermediate nanogroove width, $w_z$ = 100 nm, which provides a good balance between extraction efficiencies ($\beta$ = 0.4) and spontaneous decay rate enhancements ($P$ = 15). The other parameters are $h$ = 20 nm (m = 0 [24]) and $w_x$ = 300 nm. The spontaneous decay rate and the $\beta$-factor spectra are represented in Fig. 4(a) and 4(b) with blue dots, while the other decay mechanisms are investigated in Fig. 4(c). For the sake of comparison, the reference values, $P'$ and $\beta'$, in the absence of the cavity, are shown with the dashed-black curves. On spectral averaging, the coupling efficiency is increased by 50%, while the peak value of $P$ is 20-fold larger than $P'$. Another remarkable result is the broadband operation. Actually, because of the smallness of the cavity-mode, large Purcell factors are obtained for a low $Q$ value and the full-width-half-maximum of the $P$ spectrum is as large as the inhomogeneous broadening of the self-assembled quantum-dot emission lines, rendering the device operation successful for virtually any self-assembled QD. Equally importantly, the $\beta$-factor remains larger than 0.4 over a 100-nm spectral interval. The latter is comparable to that recently reported for direct SPP-waveguide coupling schemes [6-9], but additionally, the present architecture supports a strong enhancement of the local density of states.

For the range of wavelengths considered, one could alternatively use silver instead of gold. We have studied the performance of the source at room temperature for the silver-permittivity values tabulated in [25]. For $w_z$ = 100 nm, the spectral behaviors of $P$ and $\beta$ (not shown) are just slightly better than those obtained for gold and shown with blue dots in Figs. 4(a) and 4(b). However due to smaller silver absorption, smaller nanogrove widths (i.e. smaller mode volumes) can be investigated with silver without deteriorating the efficiency by non-radiative damping. The red crosses in Figs. 4(a) and 4(b) shows computational results obtained for $w_z$ = 50 nm. The Purcell and $\beta$ factors at the central wavelength $\lambda_0$ = 950 nm are close to 45 and 0.45 respectively. Comparatively, for the case of gold, we calculate $P \approx 30$ and $\beta \approx 0.3$ for the same value of $w_z$. In addition, note that the SPP propagation length for silver becomes $L \sim 70$ µm with a mode area comparable to that obtained for gold.

In practice, the position of the QD will not be exactly on-axis. We have analyzed the impact of the QD misalignment in the ($x,z$) plane on the performance of the source. From the CW model and especially from the reciprocity arguments used there, it is expected that the coupling coefficient $\gamma$, which only depends on the field amplitude scattered by the fundamental TE$_{01}$ at the dot position $r_0$ (Fig. 2(c2)), weakly varies with small in-plane variations of $r_0$, except for extremely narrow groove widths $w_z$. One may also expect a stronger impact for off-axis displacements $\Delta z$ along the small dimension of the nanogroove rather than along the $x$-direction. These trends are all confirmed by the CW model predictions and fully-vectorial computational results. As shown in Figs. 4(d1) and 4(d2) obtained for $w_z$ = 100 nm, the SPP-source architecture is robustly tolerant to QD misalignments. Even for the more sensitive case $w_z$ = 50 nm ($P \approx 30$ and $\beta \approx 0.3$), other calculations (not shown) have evidenced that a 20-nm shift only results in a two-fold reductions of $P$, while $\beta$ remains nearly unchanged.

Because self-assembled dots possess two near-degenerate orthogonal emission lines, for the sake of completeness, we have also investigated the decay rate of in-plane dipoles that are polarized in the direction parallel to the slit. For a centered dipole, this orientation couples neither to the SPP channels nor to the TE$_{01}$ mode and the decay rate $\Gamma_x$ is comparable to that in the bulk GaAs ($\Gamma_x \approx \Gamma_0$). Thus, in case the polarization degeneracy is not lifted, the effective $\beta$ factor would be $\beta_{eff} \approx \beta P/(P+1)$ for an arbitrary in-plane polarization. One may conclude that only the $z$-polarized dipole, for which the Purcell factor is much larger than for the other orientation, dominates the emission of the single plasmon source. This tenacity against orientation disorder is in agreement with recent findings that the orientation-averaged SE rate is strongly influenced by the dipole orientation with the maximum emission rate [26]. Finally, let us note that for an out-of-plane ($y$-polarized) QD, which also does not couple to the TE$_{01}$ mode, the Purcell factor is almost null ($\Gamma_y \approx \Gamma_0/65$).

Rapid advances in recent years for top-down III-V nanofabrication techniques including epitaxial growth, wafer bonding, QD alignments, etc… put the present plasmon-source architecture in experimental reach. The very thin QD GaAs layer may be first grown on an AlGaAs sacrificial layer. The attached small dielectric cavity dice can be further processed on top of the epitaxial sample, either in resist or in amorphous aluminum oxide obtained by wet thermal oxidation of a tiny AlAs layer. Due to the thinness of the active layer, alignment marks can be processed in the active layer, so that after metal evaporation and removal of the sacrificial layer, a second lithographic step is used to etch the SPP-channel on the top of the wafer.

## V. CONCLUSION

In summary, a design for an efficient single-plasmon source in an integrated III-V semiconductor environment has been proposed. With propagation distances of about 15 wavelengths for a gold substrate, the design offers a good balance between efficiency, confinement, and decoherence effects. With a less lossy metal substrate such as silver, our calculations show that the Purcell and $\beta$ factors can be boosted significantly by about 50% (compared to gold) with an increased SPP propagation length of ~ 70 wavelengths. Even better performance is expected if the source is operated at cryogenic temperature since the previous predictions are obtained for metal permittivity data collected at room temperature. Thus we are confident that the performance reported in Figs. 4(a) and 4(b) for gold and silver does not represent theoretical upper bounds. Additionally, let us note that the present architecture is concerned by embedded quantum emitters, but a similar principle of operation can be retained for pure two-level quantum emitters, such as molecules [27] that may be deposited above the plasmonic channel in air. Moreover, by terminating one of the plasmon channels with a mirror or Bragg resonator, a unidirectional source is achieved. The terminated channel may act as another route to increase the coupled intensity $|\gamma|^2$, thereby increasing the Purcell factor of the source, see Eq. (1).

Finally for the sake of fairness or completeness, let us note that SPP waveguides are not the sole platform that is suitable for on-chip implementation of quantum networks. Photonic crystal cavities are indeed candidates [28]. Photonic crystal waveguides offer remarkably large β'-factors for in-plane QDs, β' > 90% over a large spectral range as theoretically predicted in [29] and confirmed in [30]. By incorporating a low-Q cavity in the waveguide (like in the present work), not only the naturally high β'-factors of the waveguide is further enhanced, but in addition the SE rate of embedded QDs is boosted. By using the Fabry-Perot model in [19] and the $n_g$ and $L_{eff}$ values in [21], we estimate that P ≈ 20 and β > 0.96 is achieved for a three-hole defect cavity [28] (the so-called "L3") with $|r_b|^2$ = 0.99 and $|r_h|^2$ = 0.5. Thus we conclude that much higher β-factors are easily achieved with the photonic-crystal platform in comparison with the present plasmonic approach. However, due to the larger mode volume, the bandwidth given by Eq. (2) is also drastically lowered (< 5 nm). Let us note that slot waveguides also present high β' values over a very large spectral range [31], but they cannot be operated with self-assembled QDs. The choice of either a plasmonic or photonic crystal platform depends on the trade-off between the Purcell enhancement and the β-factor, and it remains a challenge to find an optimal solution. Hybrid plasmonic-dielectric structures [9-10,32] may also be viable alternatives.

## ACKNOWLEDGEMENTS

The authors thank Ivan Maksymov for fruitful discussions. CHG acknowledges a fellowship award from A*STAR.

――――――――――

[1] D.K. Gramotnev, and S. I. Bozhevolnyi, "Plamonics beyond the diffraction limit," Nat. Photon. **4**, 83-91 (2010).
[2] A. Babuty, A. Bousseksou, J.P. Tetienne, I. M. Doyen, C. Sirtori, G, Beaudoin, I. Sagnes, Y. De Wilde, and R. Colombelli, "Semiconductor surface Plasmon sources," Phys. Rev. Lett. **104**, 226806 (2010).
[3] E. Altewischer, M. P. van Exter, and J. P. Woerdman, "Plasmon-assisted transmission of entangled photons," Nature **418**, 304-306 (2002).
[4] S. Fasel, F. Robin, E. Moreno, D. Erni, N. Gisin, and H. Zbinden, "Energy-time entanglement preservation in plasmon-assisted light transmission," Phys. Rev. Lett. **94**, 110501 (2005).
[5] A. L. Falk, F. H. L. Koppens, C. L. Yu, K. Kang, N. de L. Snapp, A. V. Akimov, M. Jo, M. D. Lukin, H. Park, "Near-field electrical detection of optical plasmons and single-plasmon sources," Nat. Phys. **5**, 475-479 (2009).
[6] Y. Chen, N. Gregersen, T. R. Nielsen, J. Mørk, and P. Lodahl, "Spontaneous decay of a single quantum dot coupled to a metallic slot waveguide in the presence of leaky plasmonic modes," Opt. Express **18**, 12489 (2010).
[7] A. V. Akimov, A. Mukherjee, C. L. Yu, D. E. Chang, A. S. Zibrov, P. R. Hemmer, H. Park, and M. D. Lukin, "Generation of single optical plasmons in metallic nanowires coupled to quantum dots," Nature (London) **450**, 304-306 (2007).
[8] A. Gonzalez-Tudela, D. Martion-Cano, E. Moreno, L. Martín-Moreno, C. Tejedor, and F. J. García-Vidal, "Entanglement of two qubits mediated by one-dimensional plasmonic waveguides," Phys. Rev. Lett. **106**, 020501 (2011).
[9] D. E. Chang, A. S. Sørensen, P. R. Hemmer, and M. D. Lukin, "Quantum optics with surface plasmons," Phys. Rev. Lett. **97**, 053002 (2006).
[10] D. E. Chang, A. S. Sørensen, P. R. Hemmer, and M. D. Lukin, "Strong coupling of single emitters to surface plasmons," Phys. Rev. B **76**, 035420 (2007).
[11] A. F. Koenderink, "Plasmon nanoparticle array waveguides for single photon and single plasmon sources," Nano Lett. **9**, 4228-4233 (2009).
[12] A. Imamoğlu, D. D. Awschalom, G. Burkard, D. P. Di Vincenzo, D. Loss, M. Sherwin, and A. Small, "Quantum Information Processing Using Quantum Dot Spins and Cavity QED," Phys. Rev. Lett. **83**, 4204 (1999).
[13] E. Knill, R. Laflamme and G. J. Milburn, "A scheme for efficient quantum computation with linear optics", Nature (London) **409**, 46-52 (2001).
[14] E.D. Palik, *Handbook of optical constants of solids*, Academic Press, NY, Part II (1985).
[15] G. Lecamp, J.P. Hugonin and P. Lalanne, "Theoretical and computational concepts for periodic optical waveguides," Opt. Express **15**, 11042-60 (2007).
[16] M. Besbes, J.P. Hugonin, PL, S. van Haver, O.T.A. Janssen, A.M. Nugrowati, M. Xu, S.F. Pereira, H.P. Urbach, A.S. van de Nes, P. Bienstman, G. Granet, A. Moreau, S. Helfert, M. Sukharev, T. Seideman, F. I. Baida, B. Guizal and D. Van Labeke, "Numerical analysis of a slit-groove diffraction problem", J. Eur. Opt. Soc. Rapid Publ. **2**, 07022 (2007).
[17] M.L. Andersen, S. Stobbe, A. S. Sørensen and P. Lodahl, Strongly modified plasmon–matter interaction with mesoscopic quantum emitters, Nature Phys. **7**, 215-218 (2011).
[18] A. W. Snyder and J. D. Love, *Optical Waveguide Theory*, (Chapman and Hall, New York, 1983).
[19] I. Friedler, C. Sauvan, J.P. Hugonin, P. Lalanne, J. Claudon, and J. M. Gérard, "Solid-state single photon sources: the nanowire antenna," Opt. Express **17**,

2095-2110 (2009).

[20] M. P. Nezhad, A. Simic, O. Bondarenko, B. Slutsky, A. Mizrahi, L. Feng, V. Lomakin, and Y. Fainman, "Room–temperature subwavelength metallo-dielectric lasers," Nat. Photon. **4**, 395-399 (2010).

[21] P. Lalanne, C. Sauvan, and J. P. Hugonin, "Photon confinement in photonic crystal nanocavities", Laser & Photonics Rev. **2**, 514-526 (2008).

[22] H.T. Liu, P. Lalanne, X. Yang and J.P. Hugonin, "Surface plasmon generation by subwavelength isolated objects", IEEE J. Sel. Top. Quantum Electr. **14**, 1522-1529 (2008).

[23] It is worth noting that for the case $m = 0$, this enhancement in the Purcell factor is not present. For $m = 1$, it turns out that the FP resonance condition for this higher-order mode bouncing back and forth along the cavity length ($w_x$) is well-matched for lateral groove dimensions $w_x$ = 300 nm and $w_z$ = 25 nm.

[24] In Figs. 3(d) and 3(e), best performance are obtained for the $0^{th}$-order cavities; this is easily understood with the CW model by considering the round trip attenuation $|u^2|$, which is ∼ 0.99 and 0.86 for $m = 0$ and 1, implying that the absorption losses actually lower the cavity Q for large $m$.

[25] P. B. Johnson and R. W. Christy, "Optical constants of the noble metals", Phys. Rev. B **6**, 4370-4379 (1972).

[26] W. L. Vos, A. F. Koenderink, and I. S. Nikolaev, "Orientation-dependent spontaneous emission rates of a two-level quantum emitter in any nanophotonic environment," Phys. Rev. A **80**, 053802 (2009).

[27] C. Brunel, B. Lounis, P. Tamarat and M. Orrit, "Triggered Source of Single Photons Based on Controlled Single Molecule Fluorescence," Phys. Rev. Lett. **83** 2722 (1999).

[28] D. Englund, A. Faraon, B. Zhang, Y. Yamamoto, and J. Vučkovič, "Generation and transfer of single photons on a photonic crystal chip," Opt. Express **15**, 5550-8 (2007).

[29] G. Lecamp, P. Lalanne and J.P. Hugonin, "Very large spontaneous emission beta-factors in photonic crystal waveguides," Phys. Rev. Lett. **99**, 023902 (2007).

[30] T. Lund-Handsen, S. Stobbe, B. Julsgaard, H. Thyrrestrup, T. Sünner, M. Kamp, A. Forchel and P. Lodahl, "Experimental realization of highly-efficient broadband coupling of single quantum dots to a photonic crystal waveguide", Phys. Rev. Lett. **101**, 113903 (2008).

[31] Q.M. Quan, I. Bulu and M. Loncar, "Broadband waveguide QED system on a chip," Phys. Rev. A **80**, 011810 (2009).

[32] S. Schietinger, M. Barth, T. Aichele, and O. Benson, "Plasmon-enhanced single photon emission from a nanoassembled metal-diamond hybrid structure at room temperature," Nano Lett. **9**, 1694 (2009).

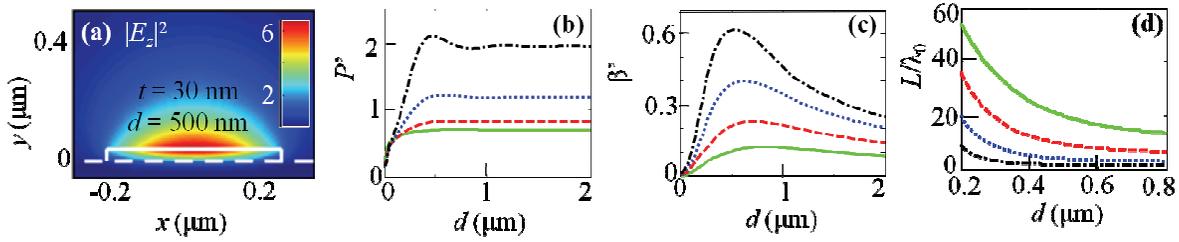

**FIG. 1 (color on line).** SPP GaAs ridge channel property. (**a**) $|E_z|^2$ of SPP mode profile for $d$ = 500 nm and $t$ = 30 nm. (**b**) Total spontaneous decay rate $P'$ of a $z$-polarized dipole located on-axis 20 nm above the gold substrate. (**c**) Corresponding $\beta'$. (**d**) SPP damping length $L$. In (**b**)-(**d**), the solid-green, dashed-red, dotted-blue, and dashed-dotted-black curves are obtained for $t$ = 25, 30, 35, and 40 nm, respectively.

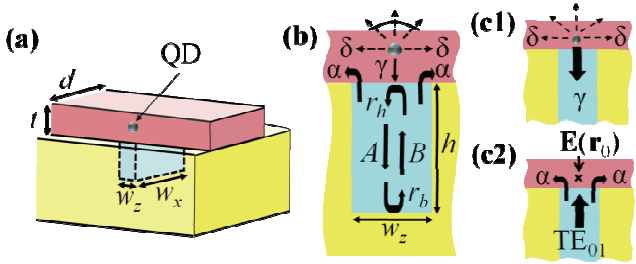

**FIG. 2 (color on line).** Semiconductor single-plasmon source and its coupled-wave (CW) model. (**a**) Sketched geometry composed of a GaAs SPP channel (magenta) on a gold (yellow) substrate with a dielectric nanogroove (blue). The QD is assumed to be linearly polarized along the $z$-direction. (**b**) The main physical quantities of the CW model are defined with the cross-sectional view. (**c1**) and (**c2**) Definition of the scattering coefficients $\gamma$ and $\alpha$ for a semi-infinite slit and illustration of the reciprocity argument to calculate $\gamma = -\mathbf{E}(\mathbf{r}_0) \cdot \mathbf{p}_z / N_T$ through the reciprocal problem shown in (**c2**). The direct coupling $\delta$ into the SPP channels and free space (dashed arrows) is neglected in the CW model.

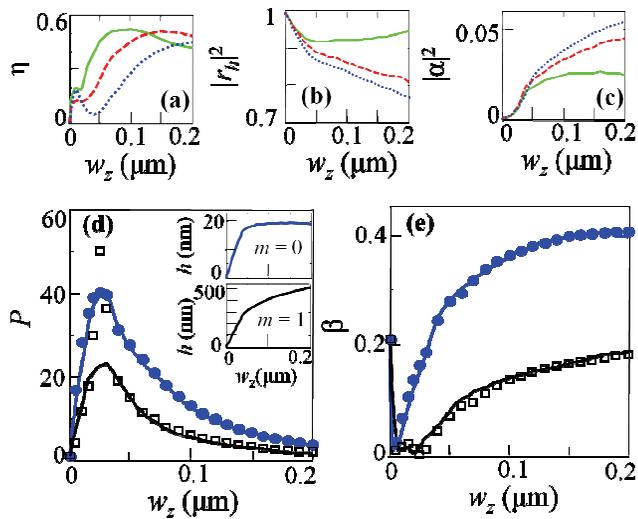

**FIG. 3 (color on line).** (**a**)-(**c**): Main quantities involved in the CW model. The solid-green, dashed-red, and dotted-blue lines are obtained for $w_x$ = 250, 300 and 350 nm, respectively. In (**a**), the vertical axis represents the fraction $\eta$ of the total spontaneous decay rate coupled to the TE$_{01}$ mode for the geometry of Fig. 2(c1). (**d**) Total spontaneous decay rate $P$. The insets show the nanogroove depths to maintain the Fabry-Perot resonance condition at $\lambda_0$ = 950 nm as $w_z$ is varied. (**e**) Corresponding $\beta$ factor. In (**d**) and (**e**), shaped markers are fully-vectorial computational results, solid curves are predictions from the CW model, and $w_x$ is taken to be 300 nm. Blue dots and black squares are for $m$ = 0 and 1, respectively. All the results are obtained for $\lambda_0$ = 950 nm.

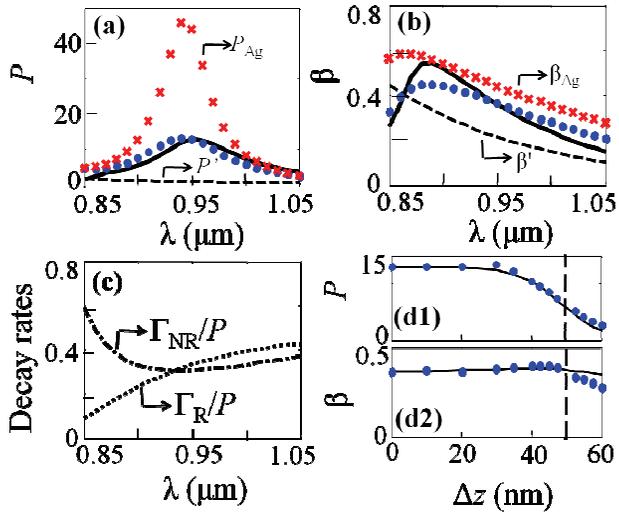

**FIG. 4 (color on line).** Performance of the semiconductor single-plasmon source. (**a**) Purcell factor for on-axis dots. Fully vectorial data (blue dots) yield results in good agreement with the CW-model (solid-black lines). (**b**) Corresponding β factor. In (**a**) and (**b**), the reference values, P' and β', in the absence of cavity, are additionally shown with dashed-black curves. For the sake of completeness, fully vectorial calculations for a cavity of dimensions $w_z$ = 50 nm, $h$ = 20 nm and $w_x$ = 300 nm in a silver substrate are shown with red crosses. (**c**), CW-predictions for the normalized spontaneous decay rates into free-space photons ($\Gamma_R/P$) and into dissipative loss ($\Gamma_{NR}/P$) (**d1**)-(**d2**) Impact of a longitudinal QD displacement $\Delta z$ (along the SPP-channel axis) on $P$ and β. The vertical blue-dashed lines indicate the position of the nanogroove boundary. All results are obtained for $w_z$ = 100 nm, $h$ = 20 nm and $w_x$ = 300 nm. Blue dots and solid black lines, like in (a) and (b).